\documentclass[aps,prl,reprint,superscriptaddress]{revtex4-1}

\usepackage{amsmath,amsfonts,amsthm,amssymb} 
\usepackage{graphicx,wasysym}
\usepackage{color}
\usepackage[colorlinks=true,linkcolor=black,citecolor=blue,urlcolor=blue]{hyperref}

\newcommand{\fr}[1]{\ref{fig:#1}}
\newcommand{\er}[1]{(\ref{eq:#1})}

\newcommand{\com}[1]{{#1}}

\begin{document} 

\preprint{DRAFT}
\title{Fast dynamics of water droplets freezing from the outside-in}

\author{Sander Wildeman}
\author{Sebastian Sterl}
\affiliation{Physics of Fluids group, University of Twente, 7500 AE Enschede, The Netherlands}
\author{Chao Sun}
\affiliation{Centre for Combustion Energy and Department of Thermal Engineering,Tsinghua University, 100084 Beijing, China}
\affiliation{Physics of Fluids group, University of Twente, 7500 AE Enschede, The Netherlands}
\author{Detlef Lohse}
\affiliation{Physics of Fluids group, University of Twente, 7500 AE Enschede, The Netherlands}
\affiliation{Max Planck Institute for Dynamics and Self-Organization, 37077, G\"ottingen, Germany}

\date{\today}

\begin{abstract}
A drop of water that freezes from the outside-in presents an intriguing problem: the expansion of water upon freezing is incompatible with the self-confinement by a rigid ice shell. Using high-speed imaging we show that this conundrum is resolved through an intermittent fracturing of the brittle ice shell and cavitation in the enclosed liquid, culminating in an explosion of the partially frozen droplet. We propose a basic model to elucidate the interplay between a steady build-up of stresses and their fast release. The model reveals that for millimetric droplets the fragment velocities upon explosion are independent of the droplet size and only depend on material properties (such as the tensile stress of the ice and the bulk modulus of water). For small (sub-millimetric) droplets, on the other hand, surface tension starts to play a role. In this regime we predict that water droplets with radii below $50\,\mu$m are unlikely to explode at all. We expect our findings to be relevant in the modeling of freezing cloud and rain droplets.
\end{abstract}

\maketitle

In mid 17th century, some peculiar tear shaped glass objects were brought to the attention of the Royal Society via Prince Rupert of the Rhine \cite{Brodsley1986}. The tears, made by dripping molten glass into cold water, were able to withstand the blow of a hammer when hit on their spherical head, while they abruptly disintegrated into fine pieces when their delicate tip was tampered with \cite{Chaudhri1998}. It was found that the toughening of the spherical head relies on the fact that glass contracts as it cools. In the cold water the outer layer of each liquid glass tear quickly cools and solidifies, encapsulating a core of molten glass. This core pulls inwards on the already solid shell as it continues to cool. Nowadays this phenomenon is used to produce toughened glass. 

What is perhaps less well known is what happens in the complementary situation, when, instead of contracting, the material of the drops \emph{expands} upon solidification. Only a handful of materials have this special property, of which water and silicon are the most common. From daily experience we know that water freezing up inside a closed container can exert extreme pressures on its container walls, as is exemplified by the fracturing of rocks by freezing water inclusions \cite{Vlahou2010} and by the playful ``ice bomb'', in which freezing water bursts out of a thick-walled metal flask \cite{Reich2016}. For \emph{unenclosed} water drops, it has been found that the combination of expansion and geometrical confinement by surface tension leads to the formation of a singular tip in the final stage of solidification~\cite{Marin2014}. However, in these experiments the spherical symmetry was purposefully broken by cooling the droplets only from one side. Symmetric, radially inwards, freezing of water droplets (as in Rupert's experiment) has been mainly studied in the context of ice formation in (rain) clouds \cite{Langham1958,Mason1960,Johnson1968,Hobbs1968,Brownscombe1968,Kuhns1968,Takahashi1970,Kolomeychuk1975}. In these studies it has been observed that the self-confinement by a rigid ice shell can cause such drops to `auto-detonate' and fly apart into pieces. This can explain (some of) the shapes of the larger ice particles in clouds \cite{Korolev2004,Rangno2005,Lawson2015} and of hailstone embryos \cite{Knight1974,Takahashi1988}. It is believed that the fragmentation of freezing water droplets can play a role in the rapid glaciation of supercooled clouds and the development of precipitation \cite{Braham1964,Chisnell1974,Rangno2005}. Ice drop bursting has also recently been observed for condensation droplets formed on chilled super-hydrophobic surfaces, where the thrown out ice fragments can speed up frost formation~\cite{Boreyko2016}. Similarly, solidification expansion and fracturing can affect the production of small silicon spheres from a melt \cite{Ueno2009}. 

Although it is clear that the fracturing of solidifying drops plays an important role in many natural and industrial processes, surprisingly little is known about the (fast) dynamics that leads up to the final cleaved particle.  With the exception of preliminary footage of Leisner et al. \cite{Leisner2014}, previous observations have been made ``after the fact'', or with a too low spatial/temporal resolution to observe the important details. \com{Qualitatively, it has been found that the success rate of a freezing drop to explode increases (1)~with the strength of the ice shell, which is influenced, for example, by the presence of dissolved gases or ions during the freezing process \cite{Langham1958,Mason1960,Brownscombe1968}, (2)~with the degree of radial symmetry of cooling \cite{Johnson1968,Hobbs1968,Brownscombe1968,Kolomeychuk1975} and (3)~with the size of the droplet \cite{Mason1960,Latham1961,Hobbs1968,Brownscombe1968,Kuhns1968}. However, as far as we know, no attempts have been made so far to model the intricate dynamics (given quantitative measures of these three factors).}


In this Letter we present high speed video footage of millimetric water droplets freezing from the outside-in, capturing both the intermittent crack formation and the final explosion in detail.  We explain the main observations by modeling both the slow freezing and stress build-up, and the fast dynamics after stress release. Finally, we discuss the implications of the models for smaller droplets, such as those found in clouds.

\begin{figure}[t]
\begin{center}\includegraphics[width=\columnwidth]{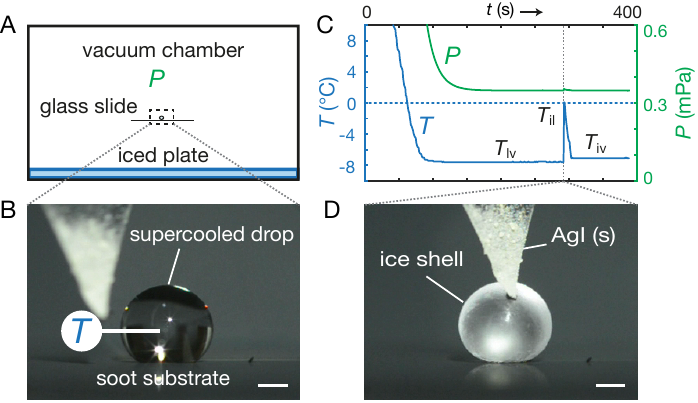}\end{center}
\caption{Experimental method. (Suppl. Video 1) \textbf{A-B}, A droplet of liquid water rests on a super-hydrophobic soot surface in the center of a small vacuum chamber. Here it becomes supercooled by evaporative cooling. An iced floor plate controls the final supercooling of the drop by providing a vapor buffer. \textbf{C}, Trace of chamber pressure $P$ and droplet temperature $T$ (measured by inserting a thermocouple in the drop in one case). The moment of ice nucleation and the three phase equilibrium temperatures involved (ice-liquid, ice-vapor, and liquid-vapor) are indicated. \textbf{D}, Once the droplet has reached a steady temperature, ice nucleation is induced by touching the drop with a tip of silver iodide (AgI). (Scale bars: 1\,mm) \label{fig:setup}}
\end{figure}

\begin{figure}
\begin{center}\includegraphics[width=\columnwidth]{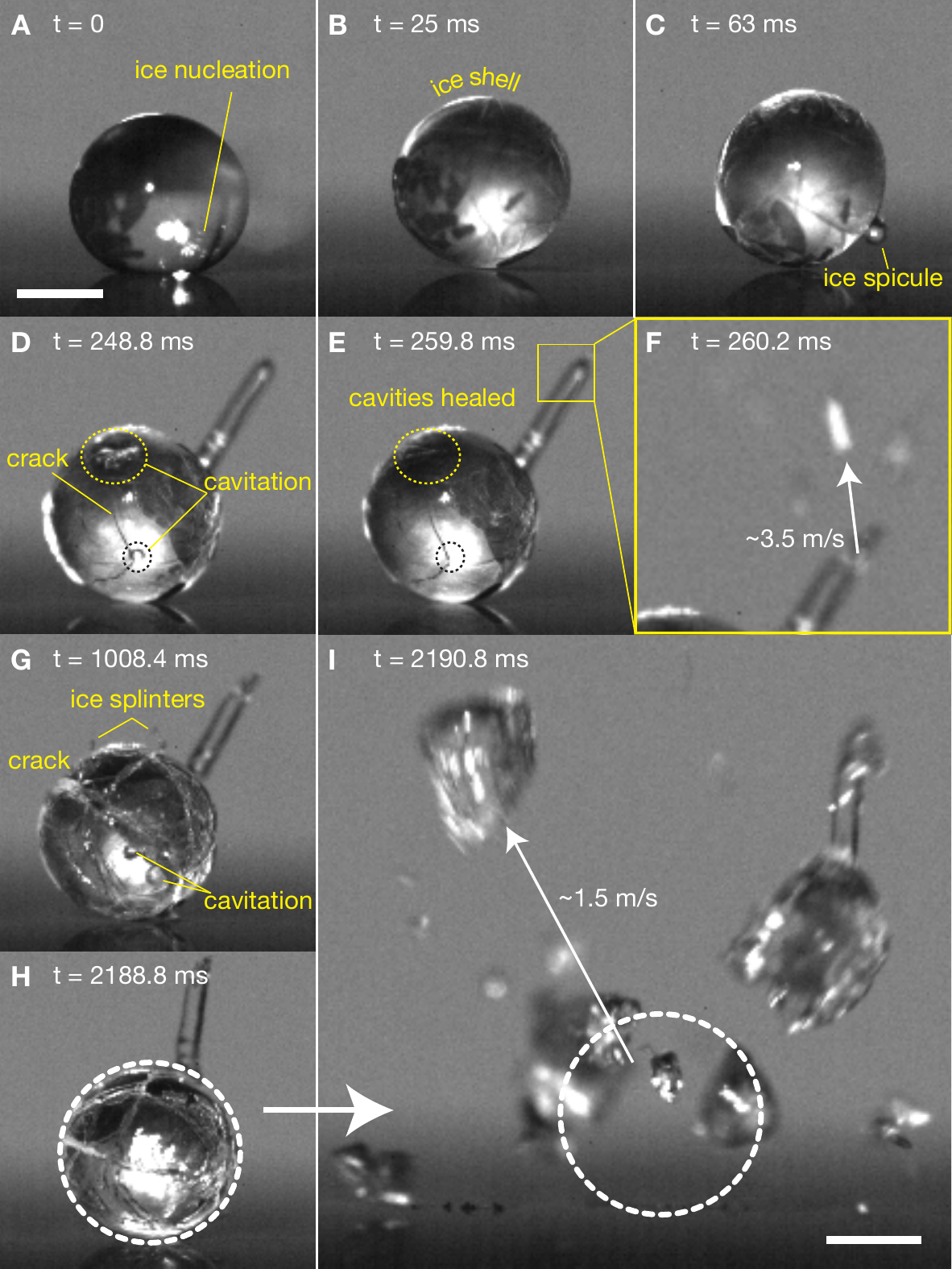}\end{center}
\caption{High speed recording of an inwards freezing water droplet. (Suppl. Video 2) \textbf{A}, Ice nucleation at the drop surface \textbf{B}, Formation of an ice shell, causing the droplet to roll over slightly. \textbf{C}, An ice spicule is pushed out from the shell. \textbf{D}, Spicule ceases to grow, the ice shell cracks and vapor cavities appear below surface (dashed circles). \textbf{E}, Cavities have completely healed. \textbf{F}, Spicule tip explodes with splinter velocities of $3.5$\,m/s. \textbf{G}, More cracks and cavities appear, accompanied by thrown out ice splinters. \textbf{H}, Droplet appearance just before the final explosion. \textbf{I}, Final explosion with fragment velocities of $1.5$\,m/s. (Scale bar: 1 mm) \label{fig:example}}
\end{figure}

In each experiment a millimetric droplet of clean, degassed water was placed on a glass slide in the center of a small vacuum chamber (Fig.~\fr{setup}A). Glass windows in the front and back sides of the chamber allowed for an unobstructed view of the droplet  (Fig.~\fr{setup}B). To ensure that the droplet kept its spherical shape, the glass surface was made super-hydrophobic by covering it with a layer of candle soot \cite{Deng2012}. Deposition of a fresh layer of soot before each trial effectively isolated the droplet from any unintended ice nuclei on the substrate (the soot itself remained inactive as nucleus down to droplet temperatures of approximately $-15^\circ$C). When the chamber is evacuated, a water droplet here rapidly cools to sub-freezing temperatures by evaporative cooling and, if nothing is done to prevent it, would spontaneously freeze \cite{Sellberg2014,Schutzius2015}. However, unrestricted evaporative cooling gives very little control over the temperature profile in the drop during the experiment. To control the droplet temperature, the floor of the vacuum chamber was covered with a layer of ice held at a pre-set temperature of $T_{iv}$ by a cooling circuit embedded in the floor of the chamber. As shown in Fig.~\fr{setup}C, this ice layer provides a buffer which (by sublimation) keeps the water vapor pressure surrounding the droplet constant at the vapor pressure $P_v(T_{iv})$ of the ice layer. The steady state temperature of the supercooled droplet is in turn determined by the equilibrium between this buffer pressure and the vapor pressure of liquid water, which occurs at a slightly lower temperature $T_{lv}$. Once a constant pressure of $340\pm 10\,$Pa (corresponding to an ice-vapor equilibrium temperature of $T_{iv} = -7.0\pm0.3\,^\circ$C) was reached in the chamber, we let each droplet equilibrate for at least three minutes before inducing ice nucleation by touching the droplet with a tip of silver iodide (Fig.~\fr{setup}D). \com{Silver iodide is one of the few materials (besides ice) which readily induces ice nucleation at low supercooling. This is attributed to the salt's ice-like crystal structure \cite{Vonnegut1947}}. After nucleation, a small fraction of the water rapidly freezes, forming thin ice dendrites which penetrate the droplet and give it a hazy appearance \cite{Mason1960,Johnson1968}. Correspondingly, the droplet temperature shoots up to the ice-liquid coexistence temperature $T_{il} = 0\,^\circ$C (cf. Fig.~\fr{setup}C). \com{Often this initial release of latent heat is accompanied by a slight roll-over or displacement of the droplet, possibly caused by a temporary (asymmetric) increase in vapor pressure at the drop's surface \cite{Schutzius2015}.} Upon further evaporative cooling a solid ice shell forms which slowly thickens (see Suppl. Mat. (SM) \S I for a detailed calculation of the slow thickening of the ice shell with time).

Figure \fr{example} shows in detail what happens after a supercooled drop comes in contact with an ice nucleus. In a few hundredths of a second the droplet is completely encapsulated by a solid ice shell (Fig.~\fr{example}B). Shortly after, a thin ice spicule is seen to be pushed out from the droplet, presumably at a weak spot in the shell (Fig.~\fr{example}C). This growing ice spicule is a first sign of the increase in pressure inside the droplet due to the expansion of the thickening ice shell. The spicule usually continues to grow to a length of approximately one droplet diameter. Once the spicule stops growing, pressure builds up again, this time leading to a sudden fracturing of the shell (Fig.~\fr{example}D). During crack formation, vapor cavities are clearly visible below the surface of the ice (dashed circles), indicating a sudden change from a high internal pressure to a pressure low enough for cavitation to occur. As the liquid water in the inclusion continues to freeze and expand these cavities and cracks gradually heal (Fig.~\fr{example}E). A droplet generally undergoes multiple of such fracturing and healing cycles during the whole freezing process. Often some small ice splinters are thrown off in these energetic events (Fig.~\fr{example}G). Figure \fr{example}F shows another interesting source of ice splinters: after the first healing phase the tip of the spicule explodes with fragment velocities of a few meters per second. Finally, about 2 seconds after nucleation, also the droplet as a whole explodes (Fig.~\fr{example}I). The velocities reached by the larger fragments are of the order of 1 m/s. Aside from some variations in the sequence of intermediate events, each droplet in our set of experiments showed qualitatively the same behavior and every droplet finally exploded. 

Often the ice drops were seen to split into two approximately equal halves (Fig.~\fr{model}A). Figure \fr{model}B shows high-resolution photographs of the cleaved surfaces of the two matching parts of such a droplet. Both halves display a clear core region, presumably formed by (partially expelled) water that was not yet solid at the time of the explosion. The cross-sections reveal some interesting details about the freezing history of the droplet, such as the partially healed crack emerging from the core and running over the whole length of the droplet. This basic picture inspired the quantitative model described next.

\begin{figure}
\begin{center}\includegraphics[width=\columnwidth]{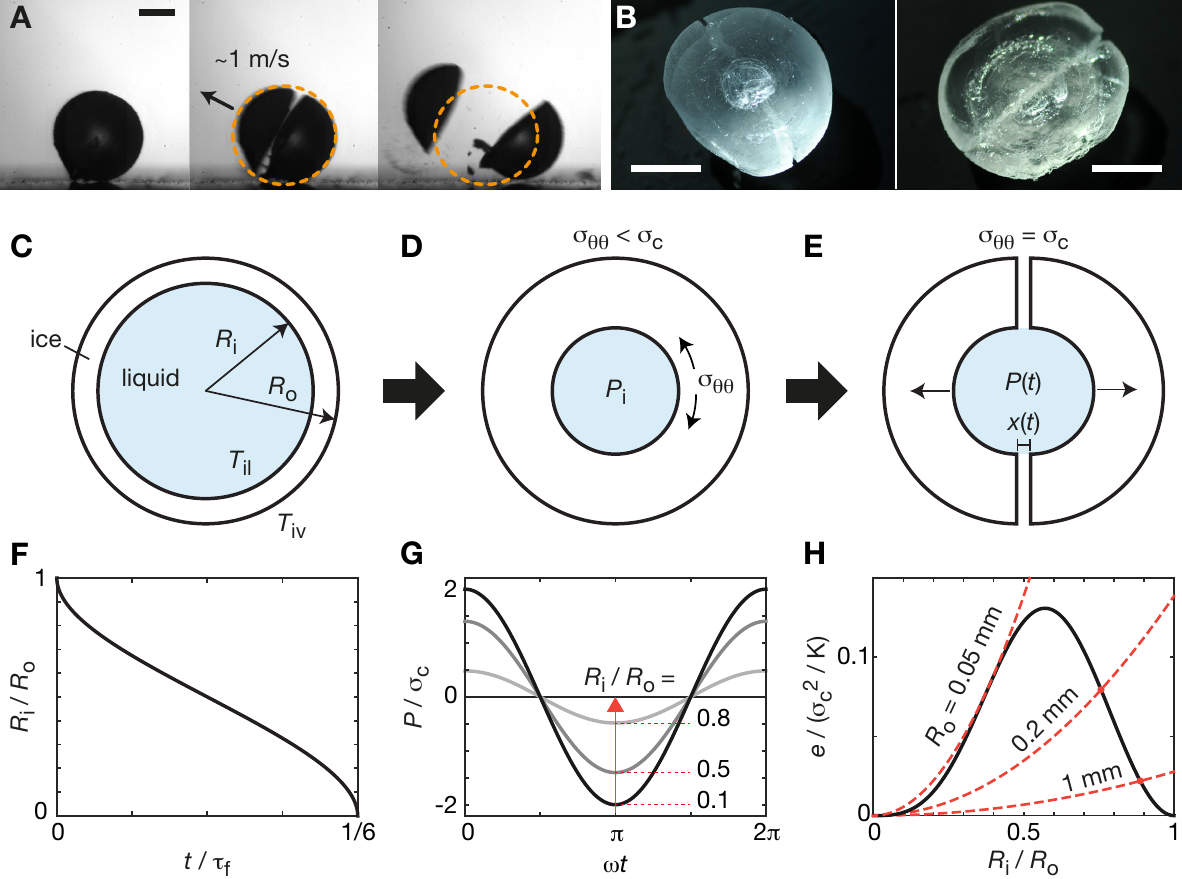}\end{center}
\caption{Basic picture of the dynamics of an inwards freezing water drop. \textbf{A}, High speed recording of a droplet bursting into two equal halves. (Scale bar: 1 mm) \textbf{B}, Cross sectional views of the halves of a burst droplet, showing a clear core region surrounded by a strong ice shell.  A healed crack emerges from the core and runs over the length of the droplet. (Scale bars: 1 mm) \textbf{C-E}, Cartoon of the freezing and bursting process. \textbf{C}, A temperature difference $T_{il}-T_{iv}$ over the ice shell makes the freezing front $R_i$ gradually proceed inward (see F). \textbf{D}, Expansion of the freezing water leads to a build-up of pressure $P_i$ in the inclusion, which in turn leads to an azimuthal stress $\sigma_{\theta\theta}$ in the ice shell. \textbf{E}, Once the stress in the ice shell exceeds $\sigma_c$ the shell cracks open. \textbf{F}, Theoretical time evolution of the freezing front, scaled by $\tau_f = (\rho L_{m} R_{o}^2)/(\kappa \Delta T)$, where $\kappa$ is the thermal conductivity of the ice shell and $L_m$ is the specific latent heat of melting (see SM \S I). \textbf{G}, Time evolution of pressure in the inclusion during the opening of a crack for three different ratios $R_i/R_o \in 0.1, 0.5, 0.8$. The pressure rapidly varies from its positive initial value to a negative value (tension) of the same magnitude. The peak-to-peak pressure increases as the relative inclusion size shrinks. \textbf{H}, Explosion energy (solid black curve) per unit volume $e = \rho v^2/2$ (scaled by $\sigma_c^2/K$) as a function of $R_{i}/R_{o}$. A maximum in energy occurs for $R_{i}/R_{o} \approx 0.57$. The dashed red lines indicate the minimum surface energy (per unit volume) required to tear apart the liquid in the inclusion for different droplet sizes. Below $R_o \approx 50\,\mu$m the required surface energy is higher for all $R_i/R_o$. \label{fig:model}}
\end{figure}

As depicted in Fig.~\fr{model}D, the expansion of the freezing water against the ice shell results in a build-up of pressure $P_{i}$ in the liquid inclusion. Because water is practically incompressible, only a small part of it has to freeze to result in a dramatic increase in the pressure. A quick calculation using the elastic bulk modulus of liquid water, $K \approx 2.2\:$GPa, and the fractional volume increase associated with the phase transition, $\beta \approx 0.09$, gives a pressure increase of $\Delta P_i \approx K \beta \, \Delta V_i / V_i \approx 2\,$MPa, when only a fraction $\Delta V_i/V_{i}$ of $1\%$ of the inclusion volume $V_i$ freezes. In reality the pressure increase would be somewhat lower due the simultaneous compression and outward expansion of the ice shell, but the order of magnitude will be the same \cite{Eshelby1959}. The internal pressure $P_i$ in turn leads to a tensile stress $\sigma_{\theta\theta}$ in the ice shell. This stress is maximal at the inner shell surface, where it is given by \cite{Landau1970}:
\begin{equation}
	\sigma_{\theta\theta} = \frac{P_{i}}{2}\left(\frac{1+2 (R_{i}/R_{o})^3}{1- (R_{i}/R_{o})^3}\right),\label{eq:icestress}
\end{equation}
where $R_i$ and $R_o$ denote, respectively, the inner and outer radius of the ice shell (see Fig.~\fr{model}C). For a thin shell of thickness $\Delta R \ll R_{o}$ this stress can be approximated as $\sigma_{\theta\theta} \approx P_{i} R_{o} / (2 \Delta R)$, while for thick shells $\sigma_{\theta\theta} \approx P_{i}/2$. In our model we will assume that once the tensile stress exceeds a critical value $\sigma_c$ associated with the tensile strength of the ice, the brittle shell will crack open. Although the tensile strength of ice does not seem to be so widely investigated, values between 0.7 and 3.1 MPa have been reported \cite{Petrovic2003}. The geometric amplification factor $R_{o}/\Delta R$ for thin shells may explain why in the initial stage liquid is easily squeezed out in the form of a spicule.

What remains to be explained is why the droplet first only shows cracks and vapor cavities, while in a later stage it completely disintegrates. For this we investigate the dynamics of the droplet just after a crack has formed. Consider the ideal situation in which a crack instantaneously completely splits the rigid shell in two equal halves (Fig.~\fr{model}E). Initially, the liquid in the inclusion will still be pressurized, exerting a force $\pi R_{i}^2 P_{i}$ on each half, acting to separate them. If the halves move apart a distance $x(t)$, the volume of the liquid inclusion increases by $dV_{i}(t) \approx \pi R_{i}^2 x(t)$ and the pressure decreases accordingly: $dP_{i}(t) = -K dV_{i}/V_i  = -(3/4)K x(t)/R_{i}$. With this, the equation of motion for each droplet half of mass $m = (2/3) \pi \rho R_{o}^3$ and moving over a distance $x(t)/2$, becomes:
\begin{equation}
  \frac{\pi}{3}\rho R_{o}^3 \frac{d^2x}{dt^2} = \pi R_{i}^2 \left(P_{i}- K\frac{3}{4} \frac{x}{R_{i}}\right),
\end{equation}
This equation represents an initially compressed mass-spring system, with the solution $x(t) = x_e (1-\cos(\omega t))$, where 
\begin{equation}
\omega = \frac 32 \sqrt{\frac{K R_{i}}{\rho R_{o}^3}}; \qquad x_{e} =\frac{4}{3} \frac{P_{i} R_{i}}{K}.\label{eq:osc}
\end{equation}
In this the initial internal pressure $P_{i}$ can be directly related to the tensile strength $\sigma_{c}$ by inverting equation \er{icestress}. For a droplet of outer radius $R_{o} = 1\,$mm, inner radius $R_{i} = 0.5\:$mm and with a shell strength of $\sigma_{c} = 3\:$MPa, one finds typical values of $P_{i} = 4.2\:$MPa, $\omega = 1.6\,$MHz and $x_{e} = 1.3\:\mu$m. The short time scale of crack opening, $2\pi/\omega = 4\,\mu$s, explains why even at the high recording rates in our experiments ($\sim 10\,$kHz) cracks seem to appear instantaneously. We note that this time is significantly shorter than the time $\Delta t \sim \mu\Delta R^2 /(x_e^2 P_i) \approx 60\,\mu s$ required for liquid water with viscosity $\mu \approx 1.8\,$mPa\,s to be pushed out into the cracks. This supports our neglect of flow in the above considerations. 

The velocity reached by each half during an oscillation cycle is $v = \omega x_{e} /2$. Upon inserting equation~\er{osc} and using equation~\er{icestress} to express $P_i$ in terms of $\sigma_c$, one discovers that the fragment velocity is independent of the size of the droplet and only depends on the material properties and the ratio $R_{i}/R_{o}$. A maximum in velocity is attained for $R_{i}/R_{o} \approx 0.57$, for which $v_{m} \approx 0.5 \sqrt{\sigma_{c}^2/K \rho}$. Again assuming $\sigma_{c} \approx 3\:$MPa, gives $v_{m} \approx 1\,$m/s, which is close to the fragment velocities observed in the experiments (see e.g. Figs. \fr{example}I and \fr{model}A and also Ref. \cite{Reich2016}). We now assess whether this velocity is always high enough for the droplet to completely split.

After the shell halves reached their maximum velocity, they overshoot their equilibrium position $x_e$. This causes the pressure in the inclusion to reverse sign, putting the liquid under tension. As shown in Fig. \fr{model}G, the maximal liquid tension is small for thin shells, but quickly increases as the inclusion size shrinks. This puts a first constraint on when an ice drop can explode: $R_{i}/R_{o}$ has to be small enough to make the pressure fall below the cavitation threshold of the liquid, which for clean water, free of gas pockets, can easily be $100\,$MPa into the negative \cite{Zheng1991,Azouzi2013}. Of course, for millimetric droplets, the micron sized gap formed by the crack will be an effective nucleation site, lowering the nucleation threshold to $\Delta P \approx 2\gamma/x_e \approx 0.1\,$MPa \cite{Harvey1944}, where $\gamma \approx 75\,$mN/m is the surface tension of water at $0\,^\circ$C. However, for a small cloud droplet of say $R_{o} \sim 10\,\mu$m, equation \er{osc} predicts a hundred fold thinner gap, and a corresponding higher threshold tension. For such droplets the cavitation threshold would form a serious barrier to explosion. A second constraint is obtained if we compare the total elastic energy released when a crack is formed, $E_{e} = 2\times (m v^2/2) = (1/2) V_{i} P_{i}^2 /K$, to the minimum surface energy, $E_\gamma = 2 \pi \gamma R_{i}^2$, required to completely split the liquid inclusion after cavitation. As shown in Fig. \fr{model}H, this puts another upper limit on the ratio $R_{i}/R_{o}$. This explains why initially, when the shell is relatively thin, droplets crack and cavitate without exploding. \com{In SM \S II we show how the above explosion criteria can be combined with the freezing model (SM \S I) to estimate the time to explosion in our experiment.} The energy comparison in Fig.~\fr{model}H also shows that for droplets with outer radii $R_{o}$ below approximately $50\:\mu$m (and $\sigma_{c} \approx 3\:$MPa) the energy required to create the new surface area is \emph{always} higher than the released elastic energy. In our model these droplets cannot explode at all. This is consistent with previous lab experiments \cite{Mason1960,Latham1961,Hobbs1968,Brownscombe1968,Kuhns1968} in which a sharp decrease in the drop bursting probability is observed around this droplet size. Also field observations on natural clouds indicate that there exists such a size threshold \cite{Braham1964,Hobbs1985,Rangno2005}.

To summarize, we have shown that millimetric water droplets that freeze radially inwards undergo a sequence of intermittent fracturing and healing events, culminating in a final explosion with fragment velocities of the order of 1 m/s. By modeling the elastic stresses in the ice shell and the fast dynamics following its failure, we have unveiled the important competing energies involved in this phenomenon. On the one hand the released elastic energy, $E_e \sim \sigma_c^2 R^3 / K$, drives the fragments apart, while on the other hand the surface energy, $E_\gamma \sim \gamma R^2$, glues them together. For millimetric droplets the elastic energy dominates, leading to the observed fragment velocities of the order of $v\sim\sqrt{\sigma_c^2/K\rho}\sim 1\,$m/s. Since only material properties appear in this expression, measuring fragment velocities of exploding water drops may provide a convenient way to probe the tensile strength of ice under different circumstances (for example in the presence of dissolved gases or ions). Below a critical radius, $R^* \sim K \gamma / \sigma_c^2 \sim 50\,\mu$m, surface energy is found to dominate. Our model predicts that below this size, ice drop explosions become impossible. This finding may be important in understanding the behavior of freezing rain and cloud droplets, which often exist in this critical size range. Although there are some indications of such a threshold in the current literature, a systematic experimental investigation into this may bear fruit. It will also be interesting to study in more detail the observed spicule formation and explosion, and the shedding of ice splinters during crack formation. Even in the absence of a final explosion these phenomena could play a role in propagating ice formation throughout a collection of supercooled droplets (cf. Suppl. Video~3).

\begin{acknowledgments}
We thank D. van der Meer and J. Snoeijer for stimulating discussions. This work was funded through the NWO Spinoza programme.
\end{acknowledgments}

%

\end{document}


\preprint{DRAFT}
\title{{\small Supplementary Material for}\\Fast dynamics of water droplets freezing from the outside-in}

\author{Sander Wildeman}
\author{Sebastian Sterl}
\affiliation{Physics of Fluids group, University of Twente, 7500 AE Enschede, The Netherlands}
\author{Chao Sun}
\affiliation{Centre for Combustion Energy and Department of Thermal Engineering,Tsinghua University, 100084 Beijing, China}
\affiliation{Physics of Fluids group, University of Twente, 7500 AE Enschede, The Netherlands}
\author{Detlef Lohse}
\affiliation{Physics of Fluids group, University of Twente, 7500 AE Enschede, The Netherlands}
\affiliation{Max Planck Institute for Dynamics and Self-Organization, 37077, G\"ottingen, Germany}

\date{\today}

\maketitle

\section*{I. Stefan problem of a radially inwards freezing water drop}

To estimate the thickness of the ice shell at a particular time $t$ after nucleation, we solve the heat transfer problem of a water droplet that slowly freezes radially inwards. Consider a spherical droplet of outer radius $R_{o}$ which at time~$t$ has reached an ice shell of thickness $R_{o}-R_{i}(t)$, where $R_{i}(t)$ is the radius of the liquid inclusion. The quasi-steady heat flow through a spherical shell held at a temperature difference of $\Delta T = T_{i}-T_{o}$ is given by (see e.g. \cite{Bejan1993})
\begin{equation}
	q_{s} = 4 \pi k \Delta T \frac{R_{o} R_{i}(t)}{R_{o}-R_{i}(t)},
\end{equation}
where $k \approx 2\:$W/(m\,K) is the thermal conductivity of the ice shell. \com{In our experiment the estimation of $q_s$ is greatly facilitated by the fact that the temperature at both the inner and outer shell surface is determined by a well defined phase equilibrium, ice-liquid and ice-vapor, respectively. The ice-vapor buffer on the bottom of the chamber ensures that the vapor pressure in the chamber stays approximately constant during the whole experiment (even when the vacuum pump is kept on) so that $\Delta T$ is constant and equal to $\Delta T = T_{il} - T_{iv} \approx 7$K.} 

This heat flow through the shell sets the rate at which latent heat $L_m \approx 330\,$kJ/kg is released as the freezing front advances inwards ($dR_{i}/dt < 0$):
\begin{equation}
	q_{f} = -4 \pi \rho L_{m} R_{i}^2 \frac{dR_{i}}{dt},
\end{equation}
where $\rho \approx 1000\:$kg/m$^3$ is the density of water. For simplicity we neglected the 9\% decrease in density associated with the liquid to ice phase transition \com{and the slight decrease in outer radius due to evaporation (to keep $T_o = T_{iv}$ we have $dR_o/dR_i = (R_i/R_o)^2 (L_m/L_s) \approx  0.12 (R_i/R_o)^2$, where $L_s \approx 2840\,$kJ/kg is the latent heat of sublimation of ice}). Setting $q_{s} = q_{f}$, we find the following differential equation for $R_{i}(t)$:
\begin{equation}
  R_{i} \frac{dR_{i}}{dt} = -\frac{k \Delta T}{\rho L_{m}}  \frac{R_{o}}{R_{o}-R_{i}}.
\end{equation}
By introducing $\tilde R_{i} =  R_{i}/ R_o$  and $\tilde t =  t/\tau_f$, with the timescale
\begin{equation}
\tau_{f} = \frac{\rho L_{m} R_o^2}{k \Delta T},
\end{equation}
this expression can be recast in an universal dimensionless form as:
\begin{equation}
  \tilde R_{i} \frac{d\tilde R_{i}}{d\tilde t} = -\frac{1}{1-\tilde R_{i}}.\label{eq:icegrow}
\end{equation}
Using separation of variables and integrating once the solution to equation \er{icegrow} is readily obtained as:
\begin{equation}
\frac{1}{3}\tilde R_{i}^3 - \frac{1}{2}\tilde R_{i}^2 + \frac 16 = \tilde t.
\end{equation}
It is clear from this relation that the droplet will be completely frozen ($\tilde R_{i} = 0$) at $\tilde t = 1/6$.

\section*{II. Estimation of the time to explosion}

The model for the thickening of the ice shell can be combined with the criteria for explosion (as derived in the main text) to obtain an estimate of the time it takes for droplet to explode.  A lower bound on the explosion time can be obtained by (numerically) finding the ratio $R_i/R_o$ for which the stored elastic energy, $E_{e} = (1/2) V_{i} P_{i}^2 /K$, at the moment of crack formation exceeds the energy, $E_\gamma = 2 \pi \gamma R_{i}^2$, required to create the fresh liquid-vapor interface (for a given droplet size $R_o$). This minimum shell thickness then serves as input for the freezing model to obtain the corresponding minimal time to explosion. An upper bound is given by the time $t = \tau_f/6$ for the droplet to freeze completely. 

In Fig. \fr{prediction} the results of this calculation are compared to experimental explosion times. Although the precise explosion time for each drop is random in nature, it is clear that the droplets did indeed always explode within the predicted time frame. In fact, with a few exceptions the explosions occur before or near the time for which the stored elastic energy reaches an optimum, i.e. when $R_i/R_o \approx 0.57$ (dashed line in Fig.~\fr{prediction}).

\begin{figure}
\begin{center}\includegraphics[width=.7\columnwidth]{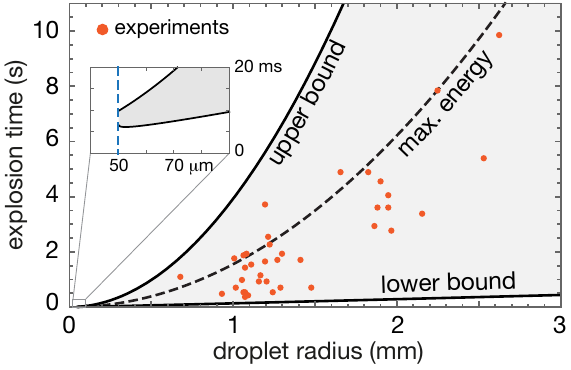}\end{center}
\caption{Explosion time as a function of the outer droplet radius $R_{o}$. The upper solid line shows the theoretical upper bound given by the total freezing time $\tau_f/6$. The lower solid line represents the theoretical lower bound given by the time it takes for the shell to reach a thickness for which sufficient internal pressure can build up. The dashed line indicates the time at which the bursting is expected to be the most energetic. Orange dots represent our experimental observations. The inset shows a zoomed in-region near the origin, highlighting the predicted inhibition of ice drop explosions below a critical radius of about 50 micron (vertical dashed line).  \label{fig:prediction}}
\end{figure}

\bibliography{refs}